\documentclass[preprint]{emulateapj}

\shorttitle{VLT detection of a red supergiant progenitor of SN 2008bk}
\shortauthors{Mattila et al.}

\def\kms{\ifmmode{\rm km\,s^{-1}}\else\hbox{$\rm km\,s^{-1}$}\fi}

\begin{document}


\title{VLT detection of a red supergiant progenitor of the type IIP supernova 2008bk}


\author{S. Mattila\altaffilmark{1}, S. J. Smartt\altaffilmark{2}, J. J. Eldridge\altaffilmark{3},
J. R. Maund\altaffilmark{4}, R. M. Crockett\altaffilmark{2}, I. J. Danziger\altaffilmark{5}}
\altaffiltext{1}{Tuorla Observatory, Department of Physics and Astronomy, University of Turku, V\"ais\"al\"antie 20, FI-21500 Piikki\"o, Finland}
\altaffiltext{2}{Astrophysics Research Centre, School of Mathematics and Physics, Queen's University Belfast, BT7 1NN, UK}
\altaffiltext{3}{Institute of Astronomy, The Observatories, University of Cambridge, Madingly Road, Cambridge CB3 0HA, UK}
\altaffiltext{4}{Dark Cosmology Centre, Niels Bohr Institute, University of Copenhagen, Juliane Maries Vej 30, 2100 Copenhagen 0, Denmark.}
\altaffiltext{5}{INAF, Osservatorio Astronomico di Trieste and Dipartimento di Astronomia-Universita di Trieste, 
via G. B. Tiepolo 11, 34131 Trieste, Italy.}




\begin{abstract}
We report the identification of a source coincident with the position of the nearby type II-P supernova (SN) 2008bk
in high quality optical and near-infrared pre-explosion images from the ESO Very Large Telescope (VLT). The SN position 
in the optical and near-infrared pre-explosion images is identified to within about $\pm$70 and $\pm$40 mas, respectively, 
using post-explosion $K_{\rm s}$-band images obtained with the NAOS CONICA adaptive optics system on the VLT. The pre-explosion
source detected in four different bands is precisely coincident with SN 2008bk and is consistent with being dominated by
a single point source. We determine the nature of the point source using the STARS stellar evolutionary models and find that its colours 
and luminosity are consistent with the source being a red supergiant progenitor of SN 2008bk with an initial mass of 
8.5$\pm$1.0 M$_{\odot}$. 
\end{abstract}
\keywords{stars: evolution -- supernovae: general -- supernovae: individual(\objectname{SN 2008bk})}

\section{Introduction}
The red supergiant progenitors of several type II-P supernovae (SNe)
have now been directly identified in pre-explosion observations. All
these are moderate mass red supergiants in the range
$\sim$7-16 M$_{\odot}$ (Smartt et al. 2004; Van Dyk, Li \& Filippenko
2003; Maund \& Smartt 2005; Maund, Smartt \& Danziger 2005; Hendry et
al. 2006; Li et al. 2006, Li et al. 2007). Recently, Smartt et al. (2008) 
have presented a volume limited systematic study of the progenitors of 20 
type II-P SNe with high-quality pre-explosion images available. They find a 
minimum initial mass of 8.5$^{+1}_{-1.5}$ M$_{\odot}$ for the progenitors and
suggest that there is a 'red supergiant problem' with the red supergiant
progenitors more massive than $\sim$17 M$_{\odot}$ remaining
undetected.

In this letter we report the identification of the progenitor of the type II-P event, SN 2008bk, making use
of adaptive optics (AO) assisted target of opportunity (ToO) observations of the SN using the ESO Very Large Telescope (VLT).
SN 2008bk was discovered by Monard (2008) on 2008 Mar. 25.14 UT in a nearby (3.9 Mpc, Karachentsev et al. 2003) Scd-type 
galaxy NGC 7793. Li et al. (2008) obtained a more precise position for the SN which is 9".2 east 
and 126".4 north of the host galaxy nucleus. It was spectroscopically classified by Morrell and Stritzinger (2008) as a type 
II-P similar to SN 1999em at 36 days after explosion on 2008 Apr. 12.4 UT. This classification is also supported by amateur
photometry$\footnote{http://www.astrosurf.com/snweb2/2008/08bk/08bkMeas.htm}$ showing a very flat plateau, consistent 
with a normal type II-P SN. NGC 7793 has a wealth of pre-discovery images
available and based on their astrometry, Li et al. (2008) identified a possible progenitor star in an archival $I$-band image from 
VLT/FORS. Subsequently Maoz and Mannucci (2008) estimated the $J$ and $K_{\rm s}$ magnitudes for the progenitor from archival near 
infra-red (NIR) pre-explosion images from VLT/ISAAC. Using accurate relative astrometry between our high-resolution 
post-explosion AO images and the pre-explosion images from the VLT we show that the possible progenitor identified by Li 
et al. and Mannucci \& Maoz is precisely coincident with the position of SN 2008bk and characterise its properties using
stellar evolutionary models.

\section{Observations and data analysis}
\subsection{Observations and data reductions}
The pre- and post-explosion observations of SN 2008bk analysed in this study are summarised in Table 1.
We obtained high quality pre-explosion imaging of the \object{SN 2008bk} site from the ESO Science Archive. The optical
observations were taken with FORS1 (0.20\arcsec/pixel) on UT3 and the NIR observations with ISAAC 
(0.148\arcsec/pixel) on UT1 and HAWK-I (0.1064\arcsec/pixel) on UT4 of the VLT. The optical frames were bias-subtracted 
and flat fielded in IRAF. Zero point magnitudes
were obtained using standard stars in the fields of Mark A, SA 110-362 and PG 1657+078 (Landolt 1992)
observed during the same night as the site of SN 2008bk. For this we adopted average colour terms and extinction
coefficients from Patat (2003). The ISAAC and HAWK-I frames were sky subtracted using sky frames created from the on-source
exposures with the IRAF XDIMSUM package, de-dithered using centroid coordinates of a bright field star visible in all 
the frames and median-combined. For the ISAAC $J$ and $K_{\rm s}$-band images zero point magnitudes were obtained using the standards FS1,
FS6, FS10, FS32 and FS114 (Leggett et al. 2006) observed before and after the SN site on the same night of observation. Average 
ESO extinction coefficients were adopted and no colour term corrections were applied. For both FORS1 and ISAAC data the average
of the zero points obtained from the different standard fields was adopted with their standard deviation as the uncertainty.
The calibration was also checked against three bright 2MASS stars which gave the same results within 0.05 mag. The SN site 
was only covered in two jittered HAWK-I $H$-band frames which were combined together. For the HAWK-I image the zero point
magnitude and its uncertainty were obtained using five 2MASS stars within the image field.

\label{obs}
\begin{table}
\caption{Pre- and post-explosion observations of SN 2008bk site.}
\begin{tabular}{lllll}
\hline\hline
Date (UT) & Instrument & Filter  & Exp. Time & FWHM \\
\hline
\multicolumn{4}{c}{{\bf Pre-explosion}} \\
2001 Sept 16.0 & FORS1   & $B$       & 300s       & 1.2\arcsec \\
2001 Sept 16.0 & FORS1   & $V$       & 300s       & 1.0\arcsec \\
2001 Sept 16.0 & FORS1   & $I$       & 480s       & 0.9\arcsec \\
2005 Apr 21.6  & ISAAC   & $J$       & 17$\times$60s & 0.5\arcsec \\
2005 Oct 17.1  & ISAAC   & $K_{\rm s}$ & 58$\times$60s & 0.4\arcsec \\
2007 Oct 16.1  & HAWKI   & $H$       & 2$\times$60s  & 0.8\arcsec \\\hline
\multicolumn{4}{c}{{\bf Post-explosion}} \\
2008 May 19.4 & NACO     & $K_{\rm s}$ &  20$\times$69s & 0.1\arcsec \\
\hline\hline
\end{tabular}
\end{table}

\label{obs}
\begin{table}
\caption{Astrometry of post- and pre-explosion images.}
\begin{tabular}{llll}
\hline\hline
                                & $I$ & $J$ & $K_{\rm s}$ \\
\hline 
Error in progenitor position (mas)   & 49/28 & 2/7 & 2/4 \\
Error in SN position (mas)           & 1/1 & 1/1 & 1/1 \\
Geometric transformation (mas)       & 50/57 & 40/40 & 23/24 \\
Total error (mas)                    & 70/64 & 40/41 & 23/24 \\
Difference in position (mas)         & 66/30 & 13/23 & 6/12 \\
\hline\hline
\end{tabular}\\
The total error has been obtained as a quadrature sum of the position errors and the geometric 
transformation RMS.
\vspace{+0.5cm}
\end{table}

\object{SN 2008bk} was observed with VLT/NACO (Rousset et al. 2003) using Target of Opportunity
(ToO) observations as a part of program 081.D-0279 (PI: S. Mattila) on 2008 May 19.4 (UT). The imaging was carried out 
in the $K_{\rm S}$-band with the S27 camera (0.027\arcsec/pixel) using the fixed sky offset imaging sequence. The AO
correction was performed using the visual wavefront sensor with the SN (m$_{\rm V}$ $\sim$ 13) itself as a 
natural guide star. The NACO data were reduced using IRAF. The jittered offset frames were median combined to 
form a sky frame, the sky subtracted images de-dithered making use of the centroid coordinates of the SN, and the 
de-dithered frames median combined. The final reduced image is of very high quality showing near-diffraction limited 
resolution of $\sim$0.1\arcsec ~for the SN (see Fig. 1).

\subsection{Relative astrometry}
To precisely determine the SN position on the pre-explosion images
we derived a geometric transformation between the pre- and post-explosion images.
Centroid positions of 26 and 30 point sources were measured in
the pre-explosion $J$- and $K_{\rm s}$-band frames, respectively, and in the post-explosion $K_{\rm s}$-band
frame. The IRAF GEOMAP task was used to derive a general geometric 
transformation between the frames.
The identification of a sufficient number of stars common between the pre- and 
post-explosion observations in other bands was not possible. Instead, we used centroid
positions of 19 stars to derive a general geometric transformation between the $I$ and $J$ band images. 
The $B$ and $V$ band images were then transformed to the $I$-band image with simple 'rscale' transformations 
(incl. $x$ and $y$ shifts and a common scale factor and rotation for $x$ and $y$) using centroid
positions of 10 stars common between the frames. A transformation was also derived
between the pre-explosion $H$ and $K_{\rm s}$-band images. The RMS values of the transformations
were adopted as the uncertainties (Table 2).

The average of the positions measured for the SN using four different methods (centroid, gauss 
and ofilter within the IRAF APPHOT package and PSF fitting with SNOOPY$\footnote{
SNOOPY, originally presented in Patat (1996), has been implemented in IRAF by E. Cappellaro. 
The package is based on DAOPHOT, but optimised for SN magnitude measurements.}$) was then transformed
to each pre-explosion image. A point-like source is clearly visible at the SN
location in the $I$, $J$, $H$ and $K_{\rm s}$-band pre-explosion images. In Fig. 1, subsections of 
the $I$, $J$, and $K_{\rm s}$ band pre-explosion images are shown together with the post-explosion 
image, all centered on the SN position. However, in the $B$ and $V$ bands no
source was detected at the SN position. To confirm the coincidence of the pre-explosion
source with the SN, its position was measured with the four different methods also used for 
the SN. The average of the measurements was then adopted as the source position and the
standard deviation as its uncertainty. In Table 2, the difference between
the source and SN positions in $I$, $J$ and $K_{\rm s}$-bands are compared with the total error budget 
in the relative astrometry (in $x/y$ coordinates). This confirms that the pre-explosion
source is coincident with the SN position within the 1$\sigma$ uncertainties in both optical and NIR
images.

\begin{figure*}[t]
\epsscale{1.05}
\plotone{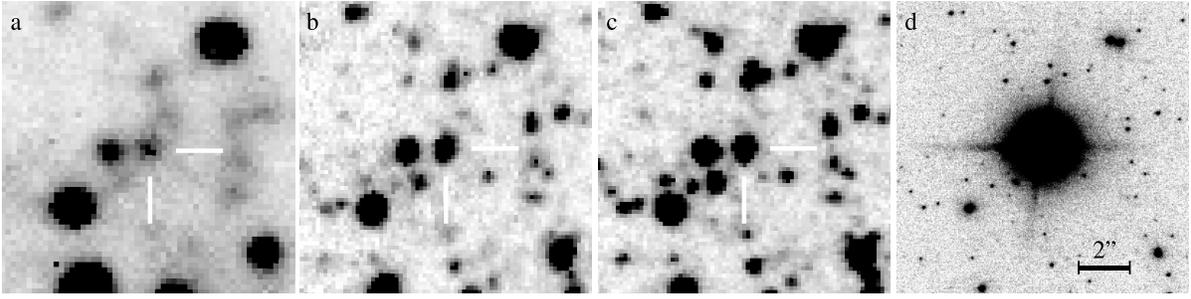}
\caption{Pre- and post-explosion images (11\arcsec $\times$ 11\arcsec) of SN 2008bk site. Each panel is centered on the SN 
position and oriented such that North is up and East is to the left. (a) Pre-explosion VLT/FORS1 $I$-band image, (b) pre-explosion 
VLT/ISAAC $J$-band image, (c) pre-explosion VLT/ISAAC $K_{\rm s}$-band image, (d) post-explosion VLT/NACO $K_{\rm s}$-band image observed with AO. 
A point-like source coincident with the SN position (marked with ticks)
is clearly visible in all the three pre-explosion frames. The $K_{\rm s}$-band source
is likely a blend between the progenitor and a fainter source $\sim$0.5\arcsec ~south making the pre-explosion source appear slightly elongated
compared to stellar PSF in this band (see Sect. 2.3).}
\end{figure*}

\subsection{Photometry in pre-explosion images}
We used the PSF fitting package SNOOPY to measure the magnitude and coordinates
for the pre-explosion source in $I$, $J$, $H$ and $K_{\rm s}$-bands. For this several suitable stars were
selected to build the model PSF. Prior to the actual PSF fitting a polynomial surface was fitted to a 
background region centered on the source position (but excluding
the innermost region around the source) and subtracted from the image. In the $I$-, $J$- and $H$-band residual images 
(with the PSF subtracted) there was little sign of the original point source, therefore confirming that the
pre-explosion object is indeed consistent with a single point source in
these bands. However, in $K_{\rm s}$-band there was a faint source left in
the residual image about 0.5\arcsec ~south of the determined SN position.
This source is at $\sim$10\% level of the original pre-explosion source peak
counts and is therefore not likely to increase significantly the
uncertainties in our $K_{\rm s}$-band photometry. The code was run both with the source position fixed (to the SN position) and 
free with very little difference in the measured
magnitudes. The uncertainties in the measurements were estimated by simulating and PSF fitting
(one at the time) nine artificial objects around the source position in the residual images.
The quadrature sum of the standard deviation of the simulated source PSF magnitudes and the uncertainty 
in the photometric zero point magnitude was adopted as the photometric error in each band. The magnitude 
corresponding to a source with a flux three times the standard deviation flux found by simulating and PSF 
fitting faint artificial objects around the SN position was adopted as the upper limit in $B$ and $I$ bands.
This yielded the following magnitudes for the pre-explosion source:  m($B$) $>$ 22.9 (3$\sigma$),
m($V$) $>$ 23.0 (3$\sigma$), m($I$) = 21.20 $\pm$ 0.19, m($J$) = 19.50 $\pm$ 0.06, m($H$) = 18.78 $\pm$ 0.11,
and m($K$) = 18.34 $\pm$ 0.07.

\begin{figure}[t]	
\epsscale{1.05}
\plotone{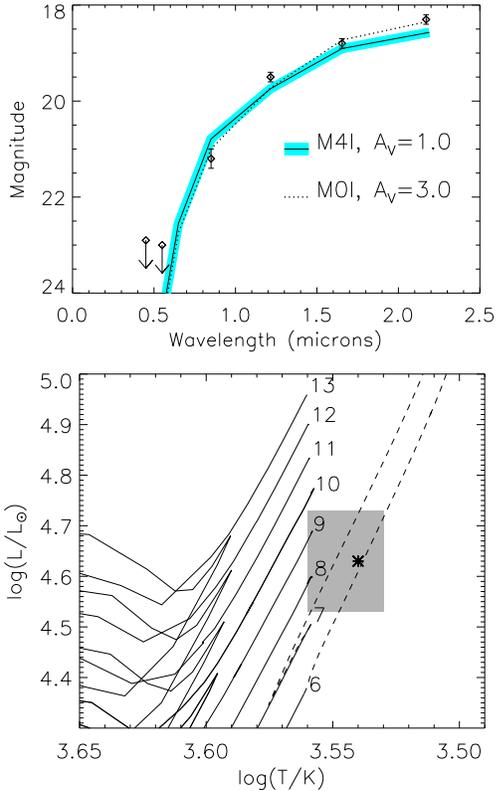}
\caption{
{\it Upper panel:} The observed SED (open diamonds) 
of the progenitor matching the reddened colours ($A_{V} = 1$) of an 
M4I star (LMC colours) from Elias et al (1985). The thick grey area
has a width of $\pm0.1^m$, to illustrate the range in observed colours
of M4I stars in each band. This band fits the observed SED, with a small
discrepancy seen in the $K$-band. An earlier M0I spectral type with a 
an extinction of $A_{V} = 3$ gives also a reasonable fit.
{\it Lower panel:} Hertzsprung-Russel (HR) diagrams showing the STARS models for initial stellar masses
between 6 and 13 M$_{\odot}$. The 6 and 7 M$_{\odot}$ model tracks during the second dredge-up
phase are indicated by dashed lines. The observed progenitor luminosity and temperature
are indicated by a star and their uncertainties as a shaded square.}
\end{figure}

\section{The progenitor of SN 2008bk}
The SN 2008bk host galaxy NGC 7793 has two similar distance estimates
of 3.91 $\pm$ 0.41 Mpc (Karachentsev et al.  2003) determined from the
tip of the red giant branch and 4.1 Mpc from the Tully Fisher relation
reported in HyperLEDA\footnote{http://leda.univ-lyon1.fr/}. Here, we
adopt the former as the more reliable estimate to be used in this
study. The metallicity at the SN 2008bk position was determined
using the relationship of Pilyugin et al. (2004) for NGC 7793.  The
offset of SN 2008bk from the host galaxy nucleus of
9.2\arcsec E and 126.4\arcsec N (Li et al. 2008) was deprojected for
the PA (83.6\degr) and inclination (53\degr) of the host galaxy as
given by HyperLEDA.  The offset corresponds to a radial distance of
3.47\arcmin~from the nucleus, at which the oxygen abundance was
determined to be about $\mathrm{12+\log(O/H)=8.2 \pm 0.1}$. Following
Smartt et al. (2008) we therefore adopt an LMC metallicity (Z = 0.008)
when estimating an initial mass for the progenitor.

Knowledge of the extinction towards the progenitor is important for
the accurate determination of the intrinsic colour, temperature and
luminosity. The foreground Galactic reddening given by Schlegel et
al. (1998) is $E(B-V)=0.019$ and there appears to be no clear evidence
that the SN suffers from high extinction. Morrell \& Stritzinger
(2008) suggest the spectrum of SN 2008bk taken on 2008 Apr 12.4 is
similar to SN 1999em (with $A_{V} = 0.31\pm0.14$, Baron et al. 2000, Smartt et al. 2003)
at +36 days after explosion. An additional handle on the total 
(Galactic+internal) extinction can be provided by determining the Balmer decrement 
for nearby H~{\sc ii} regions.  The closest H~{\sc ii} region for which a previous
spectroscopic study is available is for W13 (McCall et al. 1985), 
located 1.5\arcmin~from the site of SN 2008hk. McCall et al. (1985) provide
raw flux measurements of the Balmer lines for W13 which, assuming an
intrinsic flux ratio $H\alpha/H\beta$ of 2.85 (Hummer \& Storey 1987),
implies $c(H\beta)=0.67$. For a Cardelli et al. (1989) $R_{V}=3.1$
Galactic extinction law this implies a total extinction of
$A_{V} = 1.4$. While this estimate is clearly not directly applicable
to the line of sight to SN 2008bk, it indicates that typical 
starforming regions in the vicinity of SN 2008bk's environment 
show significant extinction.

We have used the intrinsic colours of LMC and Galactic red supergiants 
of Elias et al. (1985) to fit the observed $BVIJHK$ spectral energy
distribution of the progenitor. We find that a late-type 
M4I supergiant SED can be fit (within the uncertainties) to the
observed data with $A_{V}=1$ and  Cardelli et al. (1989) extinction law 
(Fig.\,2). The brighter $K$-band magnitude of the progenitor could be a suggestion
that one (or more) unresolved sources make up the stellar PSF. This is
also indicated by the slight elongation of the K-band pre-explosion source
(see Fig. 1) and the fact that a faint residual source was left
after subtracting the PSF (see Sect. 2.3). However,
we note that this slight discrepancy could also be due to colour differences
between the standard colours and our 2MASS-like estimate. In Fig.\,2 we also show that 
we cannot definitely rule out an earlier spectral type with higher $A_{V}$, as 
we can equally well fit the observed SED with an M0I spectrum and 
$A_{V} = 3$. Indeed one can fit even earlier types ($\sim$G-type yellow supergiants)
by invoking $A_{V}$ values up to 7. However, we favour the M4I solution with Av = 1 as 
a much higher extinction is not supported by the early SN spectra. While the SN
peak light could have evaporated circumstellar dust nearby to the SN (e.g., Dwek 1983)
therefore explaining a lower extinction towards the SN than for the progenitor we do not find
this very likely in the case of SN 2008bk. While the evaporation of a significant
amount of circumstellar dust was observed in the case of SN 2008S (Prieto et al. 2008),
there is observational evidence for a significant dusty CSM in only one type II-P event, the highly 
reddened SN 2002hh (Pozzo et al. 2006; Meikle et al. 2006). Furthermore, in the case of SN 2002hh the dusty CSM 
was found to lie well outside the dust evaporation radius resulting from an episodic mass loss 
that ceased $\sim$30 000 years before the SN explosion (Meikle et al. 2006).

The distance of 3.9$\pm$0.4 Mpc and A$_{\rm V} = 1.0\pm0.5$ results in 
M$_{\rm K}$ = -9.73 $\pm$0.26 for the progenitor. Levesque et al. (2006) show
that using M$_{\rm K}$ to determine M$_{\rm bol}$ is preferable to
using the optical bands. The best fit SED of around M4I would correspond
to $T_{eff} \simeq 3500^{+150}_{-50}$K and a bolometric correction
$BC_{\rm K} = +2.9 \pm 0.1$ (both from the Levesque et al. scale).
This results in $logL/L_{\odot} = 4.63 \pm 0.1$ (assuming $M_{\odot, bol}$ = 4.74).  
The position of the star on a HR diagram is shown in  Fig. 2, compared with model tracks from the Cambridge
stellar evolutionary code, STARS (Eggleton 1971; Pols et al. 1995;
Eldridge \& Tout 2004). This yields an initial mass estimate of $8.5\pm 1.0$ M$_{\odot}$
for the progenitor. Using the method of Smartt et al. (2008)
allowing the core-collapse to take place anytime after the end of
He-burning gives an initial mass of 9$^{+4}_{-1}$ M$_{\odot}$
consistent with the above estimate but with a larger error. We note
that adopting A$_{V}$ = 3 and an M0I spectral type would correspond
to T$_{eff}$ = 3750 $\pm$ 100 K and $logL/L_{\odot}$ = 4.8 $\pm$
0.1, which would suggest a mass of $11\pm2$M$_{\odot}$ for the progenitor.
If the $K$-band PSF magnitude is an overestimate due to 
blending, then the mass will be lower than we determined. Hence
our conclusion that this progenitor is another, moderate mass red
supergiant is unlikely to be affected.
Although the cool surface temperature of the source is also consistent with models of 
super-AGB stars (e.g. Eldridge, Mattila \& Smartt 2007), its luminosity is 
lower than would be expected for such a star.

The two best constrained progenitors for type II-P SNe before SN 2008bk were for 
SNe 2003gd and 2005cs. The progenitor for SN 2003gd was found to be a red
supergiant with a spectral type in the range of K5 to M3Ib and an initial mass of 8$^{+4}_{-2}$ 
M$_{\odot}$ (Smartt et al. 2004; Van Dyk, Li \& Filippenko 2003). The progenitor for SN 2005cs was 
found to be a red supergiant no hotter than a K5Ia type (Li et al. 2006; Maund, Smartt \& Danziger 
2005) and have an initial mass between 6 and 8 M$_{\odot}$ (Eldridge, Mattila \& Smartt 2007).

\section{Conclusions}
We have identified a source coincident with SN 2008bk in pre-explosion VLT images in four different
optical and near-IR bands making use of adaptive optics $K_{\rm s}$-band images of the SN from VLT. The colours and luminosity 
of the pre-explosion source are consistent with it being a red supergiant having an initial mass of 8.5$\pm$1.0 M$_{\odot}$. 
The coincidence of the pre-explosion source with SN 2008bk makes it the fourth intermediate mass ($\sim$8 M$_{\odot}$)
red supergiant progenitor for a type II-P SN directly detected in pre-explosion images. Our observations
also demonstrate the potential of 8m-class telescopes equipped with adaptive optics (see also Gal-Yam et al. 2005, 
Crockett et al. 2008) in precisely identifying SN progenitors in pre-explosion images.
\acknowledgements
We acknowledge funding from the EURYI scheme (SJS) and the Academy of Finland project 8120503 (SM). 
This paper is based on observations collected at the European Southern Observatory, Paranal, Chile.
We thank the referee and A. Pastorello for useful comments and the ESO staff for carrying-out the service observations.

\end{document}